\documentclass[letterpaper]{article} 
\usepackage{aaai24}  
\usepackage{times}  
\usepackage{helvet}  
\usepackage{courier}  
\usepackage[hyphens]{url}  
\usepackage{graphicx} 
\usepackage{natbib}  
\usepackage{caption} 
\usepackage{algorithm}
\usepackage{algorithmic}
\usepackage{graphicx} 
\usepackage{tabularx} 
\usepackage{multirow} 
\usepackage{amssymb} 
\usepackage{booktabs} 

\usepackage{xcolor}

\usepackage{newfloat}
\usepackage{listings}
\usepackage{booktabs} 
\usepackage{tabularx} 
\usepackage{amsmath}
\usepackage{subcaption} 
\DeclareCaptionStyle{ruled}{labelfont=normalfont,labelsep=colon,strut=off} 
\floatstyle{ruled}
\newfloat{listing}{tb}{lst}{}
\floatname{listing}{Listing}
\usepackage{lipsum}

\title{Constraint Latent Space Matters: An Anti-anomalous Waveform Transformation Solution from  Photoplethysmography to Arterial Blood Pressure}

\author {
    Cheng Bian\footnote{Corresponding author: biancheng@oppo.com},
    Xiaoyu Li, 
    Qi Bi, 
    Guangpu Zhu, 
    Jiegeng Lyu, 
    Weile Zhang, 
    Yelei Li, 
    Zijing Zeng 
}
\affiliations {
    OPPO Health Lab, Shenzhen, China\\
    \{biancheng, lixiaoyu5, zhuguangpu, lvjiegeng, zhangweile, liyelei1, zijing\}@oppo.com, 2009biqi@163.com
}

\begin{document}

\maketitle

\begin{abstract}

Arterial blood pressure (ABP) holds substantial promise for proactive cardiovascular health management. Notwithstanding its potential, the invasive nature of ABP measurements confines their utility primarily to clinical environments, limiting their applicability for continuous monitoring beyond medical facilities. The conversion of photoplethysmography (PPG) signals into ABP equivalents has garnered significant attention due to its potential in revolutionizing cardiovascular disease management. Recent strides in PPG-to-ABP prediction encompass the integration of generative and discriminative models. Despite these advances, the efficacy of these models is curtailed by the latent space shift predicament, stemming from alterations in PPG data distribution across disparate hardware and individuals, potentially leading to distorted ABP waveforms. To tackle this problem, we present an innovative solution named the Latent Space Constraint Transformer (LSCT), leveraging a quantized codebook to yield robust latent spaces by employing multiple discretizing bases. To facilitate improved reconstruction, the Correlation-boosted Attention Module (CAM) is introduced to systematically query pertinent bases on a global scale. Furthermore, to enhance expressive capacity, we propose the Multi-Spectrum Enhancement Knowledge (MSEK), which fosters local information flow within the channels of latent code and provides additional embedding for reconstruction. Through comprehensive experimentation on both publicly available datasets and a private downstream task dataset, the proposed approach demonstrates noteworthy performance enhancements compared to existing methods. Extensive ablation studies further substantiate the effectiveness of each introduced module. 

\end{abstract}

\section{Introduction}

Cardiovascular disease (CVD) is a common cause of morbidity and mortality especially among the elderly population. 
Hypertension is both a major cause and a key indicator of CVD.
Thus, it is of great importance to do continuous and regular blood pressure monitoring. 
However, the gold standard of measuring blood pressure (BP) is invasive and requires expert intervention~\cite{bp_background}. Although cuff-based BP devices have been developed, they still give rise to not only physical and psychological discomfort for patients, but also inconvenience to do longitudinal monitoring~\cite{cuffless_background_1}. Compared to conventional BP measurement, invasive arterial blood pressure (ABP) can deliver thorough insights into fluctuations in blood pressure, bearing significant essence in disease diagnosis and medical research analysis~\cite{abp_background_1}. However, due to the high requirement for sensor positioning, it is rarely utilized in clinical practice~\cite{cuffless_ref}.

Thanks to the rapid development of intelligent algorithms and smart hardware,
in recent years, photoplethysmography (PPG) has emerged as a promising alternative for measuring blood flow velocity~\cite{blood_flow_velocity}. 
In fact, as a simple, user-friendly and affordable approach, it has been widely used in devices such as smartwatches and fitness trackers. 
The key principle of PPG is to measure changes in the blood flow velocity, which correlates with the artery pressure. 
It has been revealed through some research~\cite{relate_ppg_abp, share_nature_1} that PPG and ABP signals share a common physiological factor, which can help facilitate the potential for mutual conversion.


Owing to the rapid development of deep learning techniques\cite{pan2022label, bian2021domain, bian2020uncertainty, bi2022all}, cutting-edge research in the field can be summarized into two categories, namely, estimation from blood pressure features~\cite{bp_estimation, bp_estimation_1} and waveform transformation~\cite{CycleGAN-Waveform-Reconstruction, bp-net, PPG2ABP}. 
Estimation from blood pressure features intends to predict systolic and diastolic blood pressure, but 
it inevitably discards the ABP waveform and results in the loss of cardiovascular disease details. 
It not only degrades the accuracy of blood pressure estimation~\cite{bp_estimation}, but also negatively impacts its generalization to downstream tasks such as measuring cardiac capacity~\cite{cardiac_capacity}.

\begin{figure}[!t]
    \def\svgwidth{\linewidth}
    \includegraphics[width=1.0\linewidth]{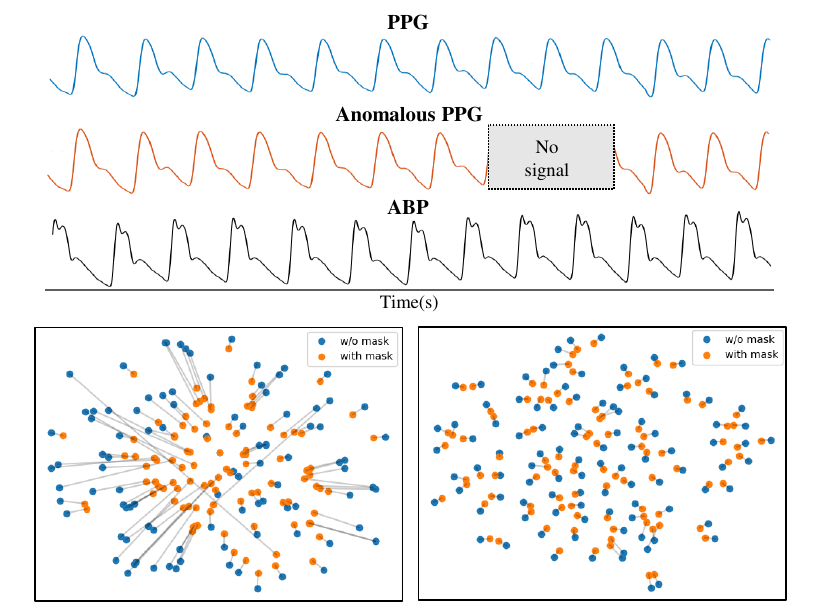}
    \caption{
    Top: Illustration of PPG, anomalous PPG and ABP signals.
    Bottom: Visualization of the latent space shift. 
    Bottom Left: Visualization (t-SNE) results from naive discriminative model (a.k.a., Swin-Transformer~\cite{swin_transformer}). 
    Bottom Right: Results from ours. Note that blue points denote the original features from PPG signals in the latent space, while orange points are the features from anomalous PPG which is random masked under the ratio of 10\%. Corresponding signals (w or w/o masks) are connected by gray lines, where the length of each line represents the distance.
    }
    \label{fig:latent_space_shift}
    
\end{figure}

To comprehensively perceive the cardiac process~\cite{bp_background_2} and measure the blood pressure from PPG, a precise mapping from the peripheral blood volume changes to the arterial blood pressure waveform is critical. 
Several studies attempted to address this bottleneck by applying deep learning-based techniques (e.g., convolutional networks~\cite{PPG-federated-learning-approach}, variational autoencoders (VAE)~\cite{auto-encoder}, Transformers~\cite{transformer_ppg}, and generative adversarial networks~\cite{CycleGAN-Waveform-Reconstruction, DAE}), as they have shown stronger representation learning capability than conventional analytical approaches~\cite{conventional_transform}. 
However, these methods highly rely on the consistency of waveform transformation. 
In practical scenarios such as wrist devices, anomalous PPG signals are captured commonly.
Consequently, such pipelines inevitably trigger a rough alignment between the decoded waveform and the ground truth, resulting in an unstable shift within the latent space as illustrated in Fig.~\ref{fig:latent_space_shift}. 
Overall, how to learn a precise PPG to ABP mapping from anomalous signal transformation remains challenging.

To address this issue, we present a novel approach named the Latent Space Constraint Transformer (LSCT) for anomalous PPG-to-ABP waveform conversion. 
In general, the codebook generated by Vector Quantized-Variational AutoEncoder (VQ-VAE)~\cite{VQVAE} can learn a relatively stable latent space by discretizing the waveform features into several codebook bases. 
This property is particularly valuable for scenarios dealing with information loss~\cite{representation_learning_1}.
However, the representation from the conventional VQ-VAE paradigm highly relies on several bases that have been discretized.
It is less expressive for our anomalous PPG-to-ABP waveform conversion task, where a certain defective basis can significantly degrade the feature representation.   

To overcome this limitation, we propose a novel Correlation-boosted Attention Module (CAM), which enables comprehensive querying of all relevant bases and allows the representation to be more stable even if some bases are defective.
It facilitates to seek a better substitution of defective basis from a global view.
Moreover, we speculate that establishing a spectrum-wise information flow can improve the representation of dependent bases derived from the codebook. 
In light of this assumption, we propose an innovative Multi-Spectrum Enhancement Knowledge~(MSEK) approach to incorporate a localized graph into the relevant bases.
Consequently, the graph embedding is able to retrieve crucial details from the spectrum (a.k.a., channel) in the decoding stage while effectively reducing the distortion from ABP construction.

Our contribution can be summarized as follows:
\begin{itemize}
\item[1)] Our work reveals the latent space shift in the defective waveform transformation, which degrades the conversion process. To address this challenge, we propose a novel Latent Space Constraint Transformer (LSCT) to quantize the latent space and reduce the distortion from anomalous PPG to ABP.

\item[2)] We propose a correlation-boosted attention module to query the highly relevant bases from a global view. It is capable of constructing comprehensive features via anomalous signals.

\item[3)] We propose a novel multi-spectrum enhancement knowledge, which treats each channel-wise feature as a node to construct a local knowledge graph flow. It boosts the expressiveness of each latent code and enhances the fidelity of the waveform conversion.

\item[4)]Through extensive experiments and analysis on MIMIC III and VitalDB, our framework can address the latent space shift well and achieves state-of-the-art performance.
\end{itemize}

\section{Related Works}

\subsection{Discriminative Waveform Transformation}

Discriminative waveform transformation approaches project the original waveform to the translating representation by leveraging either  CNN, RNN, or Transformer architectures\cite{ji2021learning, ji2023multispectral, ji2023semanticrt}. 
For CNN-based methods,
the U-Net architecture has been reported as applicable to reconstructing one-dimensional signals. For example, Athaya~\textit{et al.}~\cite{A-u-net-architecture-based-approach} utilized U-Net to reconstruct a non-invasive ABP signal from PPG signal. PPG2ABP~\cite{PPG2ABP} employed a two-stage cascaded U-Net with deep supervision to reconstruct and refine ABP waveform from PPG signal. However, the conventional convolution operations in the U-Net suffer from information loss and gradient vanishing, which undermines the representation from PPG signal~\cite{Style-transfer}.

RNN-based methods have also been extensively investigated~\cite{harfiya2021continuous}. 
However, they rely too much on hidden memory states and thus negatively impact the temporal consistency~\cite{bernard2022toward}. 
More recently, Transformer has been reported effective for this task~\cite{lan2023performer}. 
However, well-trained discriminative methods will establish a static latent space, necessitating a determined representation and thereby not readily extensible to anomalous signal processing. 

\subsection{Generative Waveform Transformation}
Generative Adversarial Networks (GANs)~\cite{GAN}~have been widely used to translate ABP waveform from PPG signal~\cite{CycleGAN-Waveform-Reconstruction}.
Recent studies step forward by using more advanced backbones for adversarial training. 
For instance, Esteban~\textit{et al.} ~\cite{RCGAN} employed Recurrent GAN (RGAN) and Recurrent Conditional GAN (RCGAN) to produce realistic multi-dimensional time-series signals. TimeGAN~\cite{TimeGAN} integrated an unsupervised GAN framework into an auto-regressive model to jointly optimize the learning of the embedding space. SigCWGAN~\cite{SigCWGAN} leveraged the path signature to capture the temporal dependencies. 
More recently, to effectively handle the longer sequences, TTS-CGAN~\cite{TTS-CGAN} introduced the Transformer model as a feature extractor. 
However, GAN models overemphasize the consistency of input and output, and thus tend to overlook the importance of the representation, which is detrimental to the task-specific waveform transformation.
Moreover, these methods also struggle with the common challenges to train GANs, such as non-convergence, mode collapse, diminished gradient and etc.

In recent years, frontier generative models like vector quantization (VQ) models and diffusion models~\cite{diffusion} have shown the potential~\cite{VQVAE}. Time-VQVAE was the first to introduce VQ-VAE into time series generation, capturing global temporal consistency~\cite{TimeVQVAE}. 

\begin{figure*}[!t]
    \def\svgwidth{\linewidth}
    \includegraphics[width=1.0\linewidth]{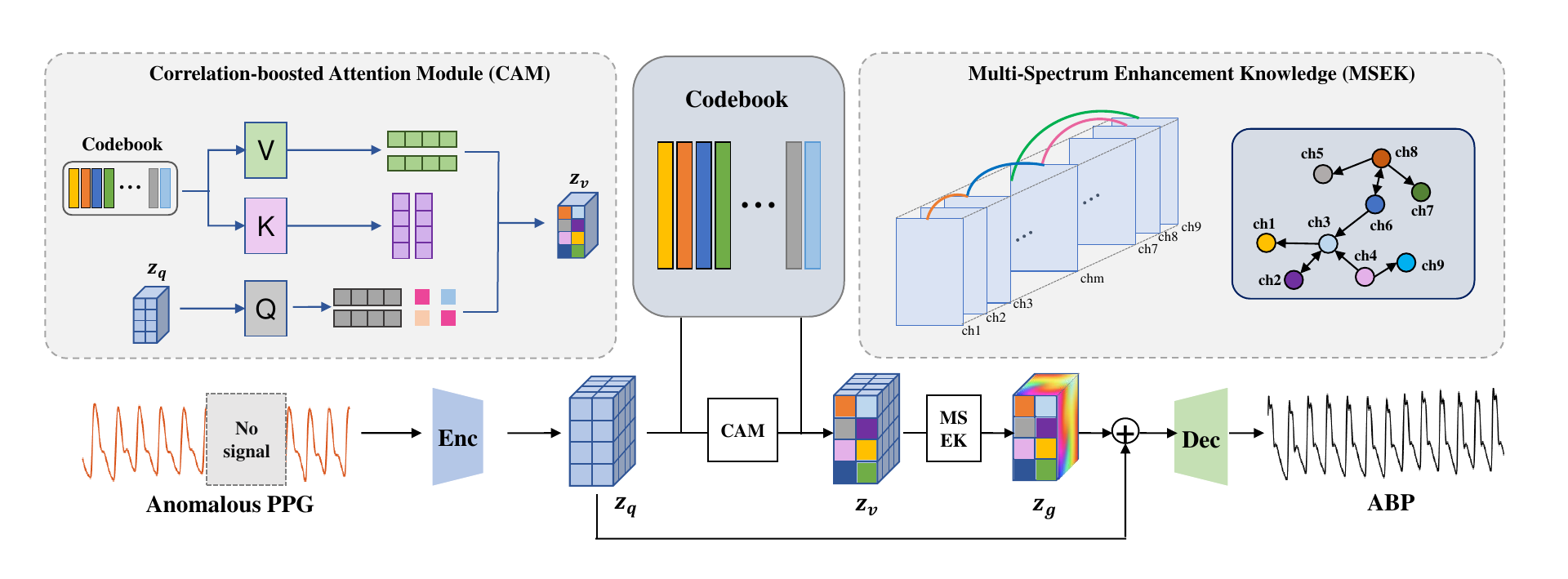}
    \caption{
        Overview of our proposed Latent Space Constraint Transformer (LSCT) framework. Firstly, anomalous PPG waveforms are used as inputs fed to the encoder. Then, the proposed Correlation-boosted Attention Module (CAM) queries relative bases to form the attentional representation $\mathbf{z_v}$. Afterward, Multi-Spectrum Enhancement Knowledge (MSEK) constructs spectrum-wise graph flow from the comprehensive representation $\mathbf{z_g}$. Finally, the ABP waveforms are decoded by the summation of the latent code $\mathbf{z_q}$ and the comprehensive representation $\mathbf{z_g}$.
    }
    \label{fig:framework}
\end{figure*}

\section{Method}
\subsection{Overview Architecture}

The overall framework of the proposed LSCT model is shown in Fig.~\ref{fig:framework}.
A codebook is served as the discretized basis dictionary to support sparse representation learning for transforming PPG waveform to ABP.

Given a sample pair of PPG and ABP ($x_p$, $\hat{x}_a$), a Short-time Fourier Transformation~(STFT) is applied to transform the pair to time-frequency domain and obtain ($u_p$, $\hat{u}_a$). The $\hat{x}_a$ and $\hat{u}_a$ serve as ground truth for the framework training, and $u_p$ is sent to the encoder as our model input.
The encoder $\mathcal{F}$ is parameterized by $\delta$, and the decoder $\mathcal{H}$ is parameterized by $\varphi$. 
The encoder compresses the signal and generates a latent code $\mathbf{z_q}$:
\begin{align*}
    \label{eq:get_z_q}
    \mathbf{z_q} = \mathcal{F}_\delta(u_p)
    \tag{1}.
\end{align*}

Then, in the CAM, the latent code $\mathbf{z_q}$ is used to query bases from the codebook to form the attentional representation $\mathbf{z_v}$. In the MSEK, spectrum of $\mathbf{z_v}$
are fused 
to produce the comprehensive representation $\mathbf{z_g}$.
The summation of $\mathbf{z_q}$ and $\mathbf{z_g}$ is then sent to the decoder to acquire the transformed ABP $\hat{u}_a$ in the time-frequency domain. The $\hat{x}_a$ in the time domain can be computed by Inverse Short-time Fourier Transform:

\begin{align*}
    \label{eq:get_u_a}
    \hat{u}_a = \mathcal{H}_\varphi(\underbrace{\textit{MSEK}(\textit{CAM}(\mathbf{z_q})}_{\mathbf{z_g}} ) + \mathbf{z_q})
    \tag{2}.
\end{align*}

The objective function is estimated to optimize all parameters by the mean square error in both the time and time-frequency domain:

\begin{align*}
    \label{eq:loss}
    \mathcal{L} = \parallel x_a - \hat{x}_a \parallel^2_2 + \parallel u_a - \hat{u}_a \parallel^2_2
    \tag{3}.
\end{align*}

\subsection{Correlation-boosted Attention Module}

Compared to typical VQ-VAE alternatives applied in computer vision tasks, PPG-to-ABP waveform transformation task requires fine-grained feature modeling. 
Thus, the core concept of CAM is to learn an attentional representation by comprehensively querying and combining all relevant discretized bases.
For this reason, we design a cross-attention mechanism for querying relative bases from the codebook. The codebook is a matrix $\mathbf{M} \in \mathcal{R}^{m\times d}$, where $m$ denotes the codebook size (i.e., number of bases) and $d$ denotes the codebook dimension (i.e., basis spectrum). The following experiment will discuss and analyze the influence of $m$ and $d$.

Note the PPG signal latent code as $\mathbf{z_q} \in \mathcal{R}^{b\times n \times d}$, $b$ denotes batch size and $n$ denotes feature map length. Cross attention is designed to query and combine different bases from the codebook, where latent code $\mathbf{z_q}$ serves as $Query$, $\mathbf{M}$ serves as $Key$ and $Value$:

\begin{align*}
    \label{eq:cross_attention}
    \mathbf{z_v} = softmax(\frac{\mathbf{z_q} \mathbf{M}^T}{\sqrt{d}}) \mathbf{M}
    \tag{4}.
\end{align*}

\subsection{Multi-Spectrum Enhancement Knowledge}

Based on the assumption that spectrum-wise information flow improves the representation of dependent bases from the codebook,
we propose the multi-spectrum enhancement knowledge module to fuse channel-wise information by graph modeling.
The graph consists of nodes and edges, $\mathcal{G} = (\mathcal{V}, \mathcal{E})$. The channels of $\mathbf{z_v}$ is treated as the nodes $\mathcal{V}=(\mathbf{h}_1, \mathbf{h}_2, \dots, \mathbf{h}_d), \mathbf{h}_i \in \mathcal{R}^{b \times n}$.
For the edges, a fully connected adjacent matrix is initialized and half of the matrix values are randomly masked with zeros. An attention coefficient can be calculated for the edge $e_{ij}$:

\begin{align*}
    \label{eq:edge_coef}
    e_{ij} = \mathbf{W} \mathbf{h}_i (\mathbf{W} \mathbf{h}_j)^T / \sqrt{n},
    \tag{5}
\end{align*}
where $\mathbf{W}$ is a learnable weight shared by all nodes.
The coefficient indicates the importance of $\mathbf{h}_j$ to $\mathbf{h}_i$. To further consider the graph structure, the coefficients between node $\mathbf{h}_i$ and its first-order neighbors $\mathcal{N}_i$ are computed and normalized:

\begin{align*}
    \label{eq:edge_attn}
    \alpha_{ij} = \frac{\exp{(e_{ij})}}{\sum_{k \in \mathcal{N}_i}\exp{(e_{ik})}} 
    \tag{6}.
\end{align*}

We can treat $\mathbf{W} \mathbf{h}_j$ as \textit{Value} and update node feature $\mathbf{h}_i$ with multi-head self-attention:

\begin{align*}
    \label{eq:update_node}
    \hat{\mathbf{h}}_i = \frac{1}{K}\sum_{k=1}^{K}\sum_{j\in \mathcal{N}_i} \alpha_{ij}^k\mathbf{W}^k h_j,
    \tag{7}
\end{align*}
where $K$ is the head number. 
Then the node features evolve as the comprehensive representation $\mathbf{z_g}$ after the graph information aggregation.

\section{Experiment}

\begin{table}[!t]
    \renewcommand\arraystretch{1.35}
    \centering
    \scalebox{1}{
    \small
    \begin{tabularx}{1\linewidth}{@{}X c c c@{}}
        \toprule[1pt]
        Item & MIMIC III & VitalDB  & OML\\
        \midrule
        
        Number of subjects & {942} & {144}  & {20}\\
        Number of segments& {127,260} & {523,080}  & {24,234}  \\
        Average SBP (mmHg) & {134.19} & {120.38}  & {120.84}\\
        SD of SBP (mmHg)& {22.93} & {19.63}  & {12.19}\\
        Average DBP (mmHg)& {66.14} & {60.21}  & {79.10}\\      
        SD of DBP (mmHg)& {11.45} & {12.60}  & {8.52}\\
        \bottomrule
    \end{tabularx}}  
    \captionsetup{width=1\linewidth}
    \caption{Summary of statistical information of the proposed datasets.}
    \label{table2}
\end{table}

\subsection{Datasets and Evaluation Metrics}

In the experiment, we evaluate our method on two public datasets and one private dataset. Particularly, public datasets including MIMIC-III and VitalDB are curated for the wave transformation task. Meanwhile, the private OML dataset is developed for the BP values regression, which is the subtask of PPG-to-ABP wave transformation used to validate the generalization of our method. Statistical details on three datasets can be referred to Table~\ref{table2}.

\paragraph{MIMIC-III.}  As one of the critical datasets, MIMIC-III is conducted by using a processed subset of the Medical Information Mart for Intensive Care~(MIMIC-III)
~\cite{MIMIC-III, MIMIC-III_2, physionet}. It encompasses a collection of 12,000 PPG, ECG, and ABP signals, which are synchronously sampled from 942 patients in practical clinical environments. All signals are sampled at 125 Hz via filtering and preprocessing. Specifically, we segment ABP signals with a length of 8.192 seconds, each slice containing 1,024 points. Likewise, we employ the same operation on the PPG signal and align it to the sampled ABP signals. Overall, 127,260 ABP and PPG segment pairs are selected for evaluation.


\paragraph{VitalDB.} The VitalDB dataset includes synchronized ECG, PPG, and ABP recordings~\cite{vitalDB}. It comprises 6,388 records from 6,090 ICU patients who underwent surgeries at the Seoul National University Hospital. This dataset provides a substantial collection of over 16,000 PPG and ECG signals, randomly selected from 550 subjects, with a duration of 8.192 seconds (i.e., 1,024 points) and sampled at a frequency of 125 Hz.



\paragraph{OML.} Our private dataset, denoted as OML, consists of 24,234 systolic blood pressure~(SBP) and diastolic blood pressure~(DBP) value pairs over 20 subjects, where SBP and DBP substitute the ABP waveform records in this dataset for longitudinal tracking efficiency. All SBP and DBP values are sampled from the high-precision blood pressure monitor, whereas PPG signals are captured by our professional wearable device. The collection period spans a duration of at least 30 days for validity purposes. We manually align an SBP-DBP value pair for one PPG record. To ensure consistency in the sampling rate and segment length across the PPG dataset, we downsample the PPG signals to 125Hz and divide them into multiple segments of equal length with 1,024 for analysis. 

\begin{table*}[!t]
    \renewcommand\arraystretch{1.35}
    \small
    \centering
    \begin{tabularx}{1.0\linewidth}
    {l >{\centering\arraybackslash}X@{}>{\centering\arraybackslash}X  >{\centering\arraybackslash}X|@{}>{\centering\arraybackslash}X@{}>{\centering\arraybackslash}X@{}>{\centering\arraybackslash}X@{}}
    \toprule
     \multirow{2}{*}{Methods}  & \multicolumn{3}{c|}{MIMIC III}  & \multicolumn{3}{c}{VitalDB} \\
     \cline{2-7}
       & RMSE~$\downarrow$ & PRD~$\downarrow$ & FD~$\downarrow$ & RMSE~$\downarrow$ & PRD~$\downarrow$ & FD~$\downarrow$  \\
    \hline
    \multicolumn{7}{l}{\textit{Discriminative Waveform Transformation}} \\
    \hline
    LSTM~\citep{harfiya2021continuous}  & {0.156} & {3.937} & {3.711} & {0.033} & {0.504} & {0.919} \\
    Unet~\citep{A-u-net-architecture-based-approach}  & {0.124} & {3.618} & {3.427} & {0.020} & {0.311} & {0.745} \\
    PPG2ABP~\cite{PPG2ABP} & {0.101} & {3.743} & {3.428} & {0.019} & {0.281} & {0.729} \\
    \hline
    \multicolumn{7}{l}{\textit{Generative Waveform Transformation}} \\
    \hline
    CycleGAN~\citep{CycleGAN-Waveform-Reconstruction} & {0.163} & {4.825} & {4.809} & {0.048} & {0.706} &  {1.520} \\
    TTS-CGAN~\citep{TTS-CGAN} & {0.150} &  {4.062} & {3.869} &{0.032} &{0.475} &{1.036} \\
    VQ-VAE~\citep{VQVAE}  & {0.108} & {3.873} & {3.693} & {0.019} & {0.285} & {0.641} \\
    Time-VQVAE~\citep{TimeVQVAE}  & {0.109} & {3.877} & {3.701} & {0.020} & {0.298} & {0.659} \\
    \hline
    Ours & \textbf{0.097} & \textbf{3.402} & \textbf{3.107} & \textbf{0.016} & \textbf{0.258} & \textbf{0.566} \\
    \bottomrule
    \end{tabularx}
    \caption{Quantitative comparisons with the state-of-the-art methods on MIMIC III dataset and VitalDB dataset. ``$\downarrow$'' indicates that metrics with lower values correspond to better performance. 
    }
    \label{table1}
\end{table*}

\paragraph{Masked Strategy.} The random masked strategy will be implemented on the PPG signal in both of the mentioned datasets. For example, a 50\% masked ratio means that a zero mask, with a size equal to half the length of the original series, will be applied to modify the PPG series.

\paragraph{Evaluation Metrics.} Root Mean Squared Error~(RMSE), Percentage Root Mean Squared Difference~(PRD) and Fr\'{e}chet Distance~(FD)~\cite{frechet_distance} are commonly utilized to measure the waveform transformation performance~\cite{tranformation_metric_1, transformation_metric_2}. We use the three metrics to compare different models and variants.
RMSE measures the stability between two signals $S$ and $\hat{S}$:

\begin{align*}
    RMSE = \sqrt{\frac{1}{N} \sum_{i=1}^N (S_i - \hat{S}_i)^2}.
    \tag{8}
\end{align*}
PRD measures the distortion between the signals:

\begin{align*}
    PRD = \sqrt{100 \times \sum_{i=1}^N(S_i-\hat{S}_i)^2 / \sum_{i=1}^N(S_i)^2}.
    \tag{9}
\end{align*}
FD distance measures the similarity between the two signals, considering the location and the order of data points:

\begin{align*}
    FD = \min_M \max_{(s, \hat{s})\in M} \parallel s - \hat{s} \parallel^2_2,
    \tag{10}
\end{align*}
where $s$ and $\hat{s}$ are the data points in the signal $S$ and $\hat{S}$ respectively, and $M$ is the matrix $\{(s, \hat{s})|s \in S, \hat{s} \in \hat{S}\}$. Besides, for the downstream regression task, we utilize mean average error~(MAE), mean error~(ME) and standard deviation~(SD) as metrics.


\subsection{Implementation Details} 


Our algorithm is implemented using the PyTorch framework~\cite{paszke2019pytorch}. The encoder $\mathcal{F}$ and decoder $\mathcal{H}$ utilized in our study are implemented with four identical blocks of Swin Transformer~\cite{swin_transformer}. The encoder blocks had channel sizes of 8, 16, 32, and 64 respectively, while the decoder blocks had channel sizes of 64, 32, 16, and 8 correspondingly. Downsample block in the encoder and upsample block in the decoder are implemented with PatchMerging~\cite{swin_transformer} and transpose convolution layer~\cite{deconvolution}. For the CAM module, the size of the codebook $m$ is set to 128, and the codebook dimension $d$ is configured as 64. Random initialization is applied. For the MSEK module, $n$ is set to 40 and $K$ is set to 8. We utilize the vanilla Adam optimizer~\cite{adam_optimizer} to update all parameters, with the learning rate of 1e-3, weight decay of 1e-5, and batch size $b$ of 512. The cosine annealing schedule~\cite{learning_rate_scheduler} is employed. 


\begin{figure*}[!t]
    \def\svgwidth{\linewidth}
    \includegraphics[width=1\linewidth]{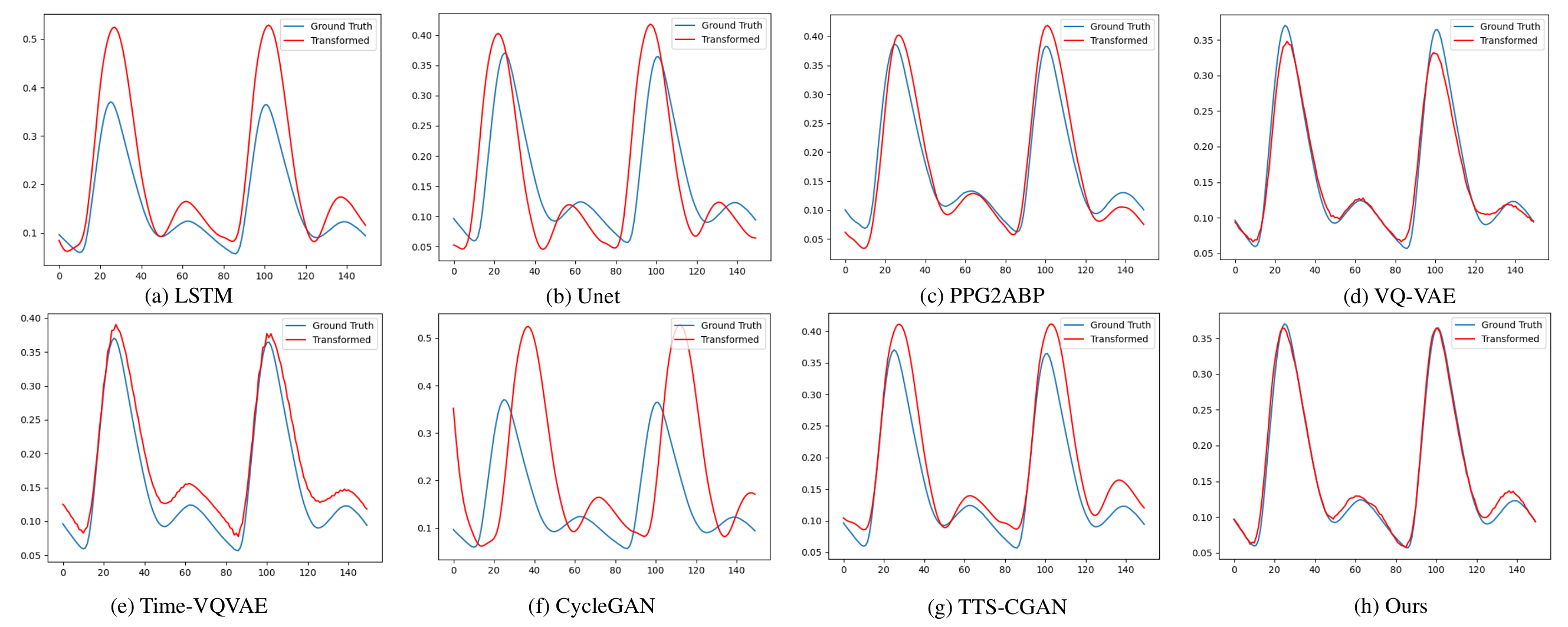}
    \caption{Qualitative comparisons of ours with the state-of-the-art methods on the MIMIC-III dataset, using the mask ratio of 10\%.}
    \label{fig3}
\end{figure*}

\subsection{Ablation Studies}
\begin{figure}[!t]
    \def\svgwidth{\linewidth}
    \begin{center}
    \includegraphics[width=1\linewidth]{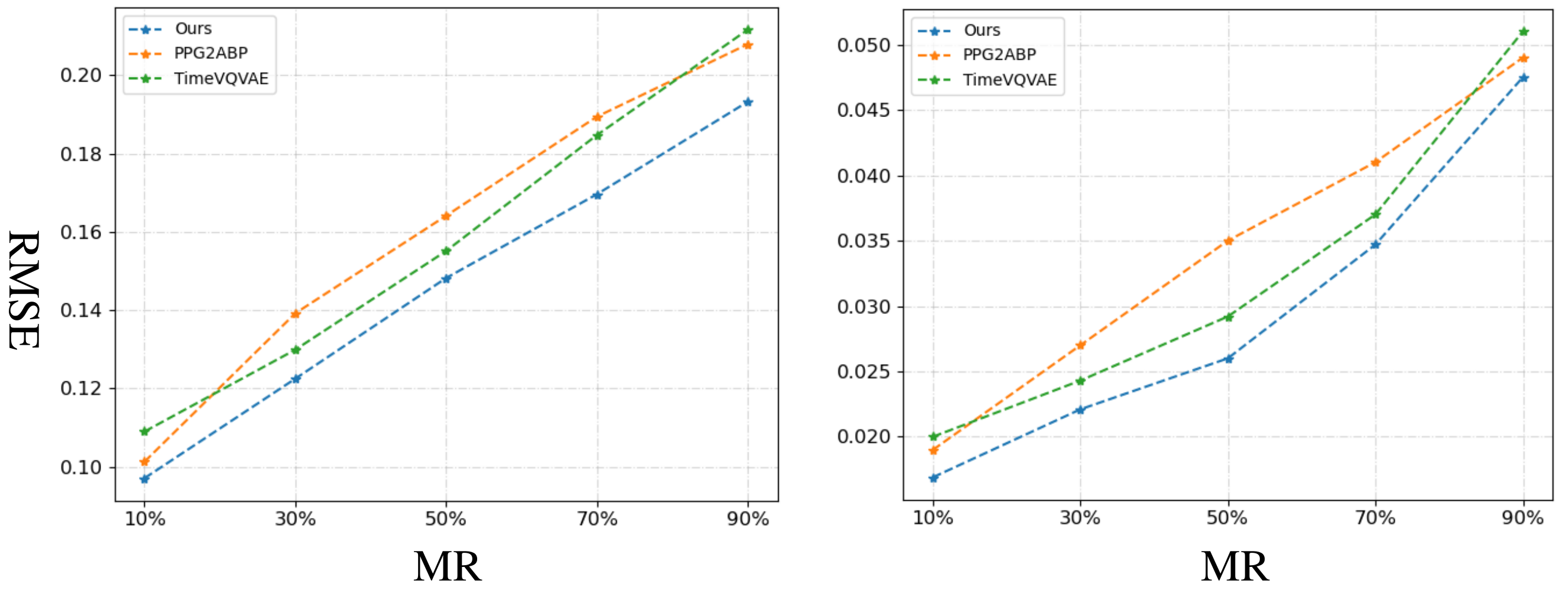}
    \end{center}
    \caption{RMSE comparison of typical state-of-the-art methods and ours given by different MRs.}
    \label{fig:comparison_diff_mr}
\end{figure}

To evaluate the effectiveness of the proposed modules individually, we conduct comprehensive ablation studies on two datasets, as listed in Table~\ref{tab:ablation_main}, in which four types of module ablations, as well as the baseline denoted as the Swin-Transformer, will be included. 
{Note that baseline is the vanilla VQ-VAE using K-nearest neighbors~\cite{knn} as the basis selection. }Other attempts include our proposed CAM and MSEK modules. As we can see, 12.5\%, 6.6\%, 7.1\% decreases have been achieved in RMSE, PRD and FD respectively when involving the CAM module, demonstrating that our method can facilitate codebook selection and query valuable basis for the waveform transformation. After that, notable overall performance gains have been observed by introducing the MSEK module, verifying that constructing graph flow in the channel can contribute to enhancing the expressiveness of the basis. Furthermore, data ablations are being performed as shown in Table~\ref{tab:ablation_main}. 
As the increasing masked ratio $MR$, the performance of baseline in RMSE, PRD and FD decreases in the average of 111.8\%, 97.7\% and 128.7\%, respectively. On the contrary, our method surpasses the baseline under to the same $MR$ by a large margin with average improvements of 101.6\%, 89.3\% and 88.3\% in RMSE, PRD and FD, suggesting that our method has the ability to deal with the anomalous signals transformation effectively and robustly.
From another perspective, Fig.~\ref{fig:comparison_diff_mr} illustrates a fair comparison of RMSE from typical state-of-the-art generative model, discriminative model and ours given by different $MR$s. It can also be noticed that our proposed method is anti-anomalous, surpassing frontier robust methods.
\begin{table}[!t]
    \renewcommand\arraystretch{1.35}
    \centering
    \small
    \setlength{\tabcolsep}{1.1mm}{
    \begin{tabularx}{1\linewidth}{@{}l  c c c c c c}
       \toprule
       Methods  & {CAM} & {MSEK} & {MR} & {RMSE} & {PRD} & {FD}\\
       \midrule
       Baseline & & &  {10\%} & {0.019}& {0.285}& {0.641}\\
       
    \midrule
       \multirow{3}{*}{Module ablations}
        &  & $\checkmark$  & {10\%} & {0.018}& {0.275}& {0.606}\\
        & $\checkmark$ &  & {10\%} & {0.017}& {0.262}& {0.575} \\
        & $\checkmark$ & $\checkmark$ & {10\%} & \textbf{0.016}& \textbf{0.258}& \textbf{0.566}\\
       \bottomrule
       \multirow{8}{*}{Data ablations}
     &  &  &  {30\%} & 0.025 & 0.342& 0.841 \\
     & $\checkmark$ & $\checkmark$  & {30\%} & \textbf{0.022}& \textbf{0.322}& \textbf{0.798} \\
     &  &  &  {50\%} & 0.030& 0.428& 0.882 \\
    & $\checkmark$ & $\checkmark$  & {50\%} & \textbf{0.026}& \textbf{0.379}& \textbf{0.833}\\
    &  &  &  {70\%} & 0.037& 0.557& 1.635 \\
    & $\checkmark$ & $\checkmark$  & {70\%} & \textbf{0.034}& \textbf{0.505}& \textbf{1.112}\\
    &  &  &  {90\%} & 0.052& 0.745& 1.903 \\
    & $\checkmark$ & $\checkmark$  & {90\%} & \textbf{0.047}& \textbf{0.692}& \textbf{1.520}\\
       \bottomrule
    \end{tabularx}}
    \caption{Ablation studies on the VitalDB dataset. Note that the mask ratio on the data ablations ranging from 10\% to 90\%.}
 \label{tab:ablation_main}
 \end{table}

\begin{table}[!t]
    \renewcommand\arraystretch{1.35}
    \centering
    \scalebox{1.0}{
    \small
    \begin{tabularx}{1\linewidth}{@{}X c c c c@{}}
        \toprule[1pt]
        Graph Flow Variants & RMSE & PRD & FD\\
        \midrule
        Self-attention (SA) & {0.127} & {4.457} & {4.129}  \\
        Graph-based Codebook (GBC)  & {0.138} & {4.562} & {4.351}  \\
        Ours  & \textbf{0.097} & \textbf{3.402} & \textbf{3.107} \\
        \bottomrule
    \end{tabularx}}
    \captionsetup{width=1.\linewidth}
    \caption{Investigation of graph flow variants of the MIMIC III datasets.}
    \label{table4}
\end{table}
\subsection{Investigation on MSEK}
To investigate which information flow can promote expressiveness of basis, we conduct an experiment on MIMIC III and analyze its impacts on different variants. Quantitative results are shown in Table~\ref{table4}. One of the compared methods is self-attention denoted as SA, which directly utilizes the attentional representation as query, key and value to compute the spectrum-wise weights on each basis. The other compared method is the graph-based codebook denoted as GBC, which constructs a codebook graph on relative bases. Compared to our method, it is noticed that the methods of SA encounter performance drops of 30.9\%, 31.0\% and 32.9\%, meanwhile GBC has the performance drops of 42.2\%,34.1\% and 40.0\% in RMSE, PRD and FD, respectively. The underlying reason is the way of self-attention employing the global graph for the computation, where not every channel of the basis is critical for the expression of the waveform transformation. Besides, GBC builds a computational graph between basis, which does harm to the orthogonality of the basis from the discretized codebook. In contrast, our methods can not only maintain the information flow within spectrum space, but also ensure the orthogonality between bases.

\subsection{Impact of the Size of Codebook and Dimension of Bases}
\begin{figure}[!t]
    \def\svgwidth{\linewidth}
    \begin{center}
    \includegraphics[width=1\linewidth]{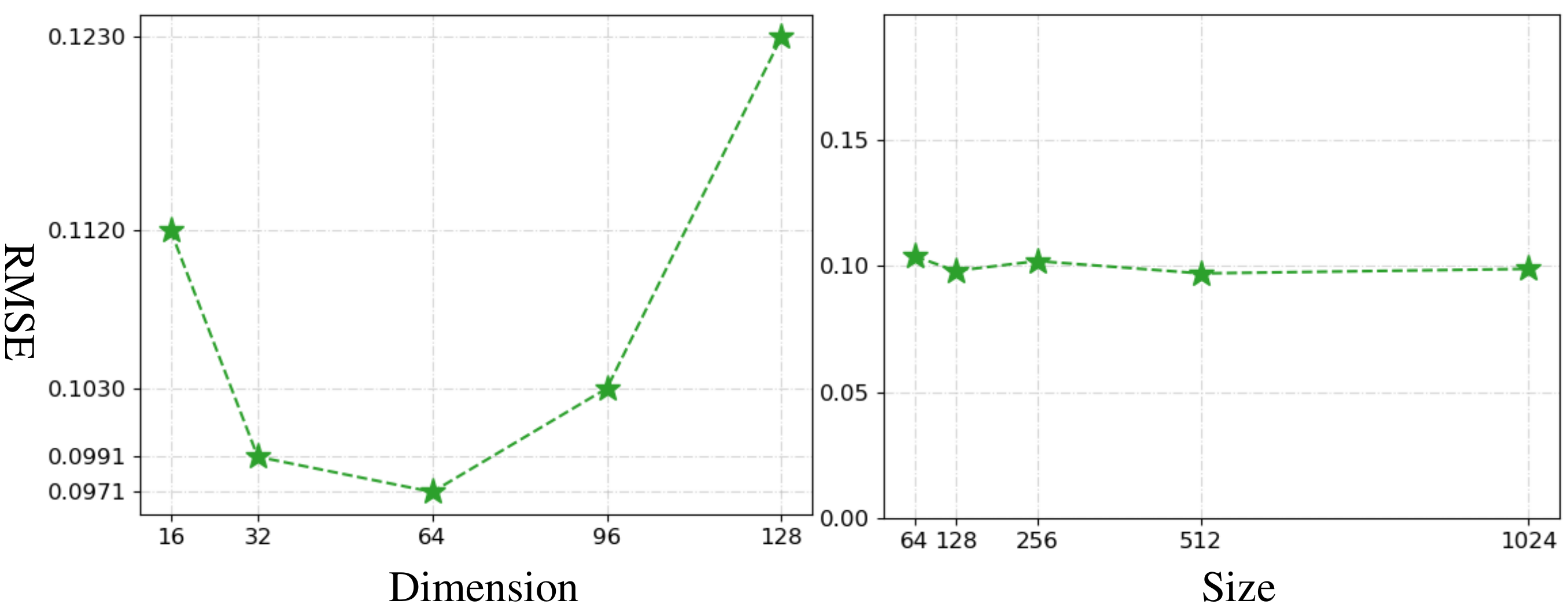}
    \end{center}
    \caption{Performance with varied size of codebook and dimension of bases.}
    \label{fig:comparison_codebook}
\end{figure}

To understand whether the size of a codebook will affect transformation performance, we employed various codebook sizes (64, 128, 256 and 512) to ascertain its effect on the transformation performance. The result is plotted in Fig.~\ref{fig:comparison_codebook}. It can be seen that variation in size settings has a negligible impact on the performance of our methodology. The dimension of bases will influence the construction of the knowledge graph flow. Therefore, another experiment is performed to track the performance manifestations under various dimensions of bases. As shown in Fig.~\ref{fig:comparison_codebook}, we achieve the best performance by the optimal dimension $d$ of 64.

\subsection{Comparison with State-of-the-arts}

To validate the efficacy of the proposed method, a comparison is made between the proposed framework and competitive methods. For a fair comparison, three representative discriminative waveform transformation methods and four generative waveform transformation methods are chosen in our experiment under the same experimental configuration and masked ratio~(10\%). Table~\ref{table1} summarizes the experimental results of the proposed method. The proposed method yields the best against all compared methods. In particular, our method has shown a performance improvement of at least 4.1\% in RMSE compared to discriminative methods, indicating that our method exhibits the capability to mitigate and overcome latent space shift ensuring robust and reliable performance. Similarly, our model's performance has surpassed that of generative methods in both criteria. This observation further corroborates that a stable latent space can effectively eliminate distortions during wave transformation.  From Fig.~\ref{fig3}(a), (b), (c) and (h), we note that the performance of PPG2ABP surpasses that of Unet and LSTM. The reason is PPG2ABP contains a cascaded Unet model, which allows a coarse-to-fine refinement procedure so that it alleviates anomalous PPG signals to some extent. As to Fig.~\ref{fig3}(d), (e), (f) and (g), we found that the overall performance of VQ-VAE alternatives is better than GANs. It substantiates our view that representation is critical to waveform transformation. Contrasting to these approaches, CAM and MSEK modules offer attentional representation and spectrum-wise information flow, thereby effectively mitigating the latent space shift and yield optimal performance.

\subsection{Generalization of Downstream Task}
As presented in Table~\ref{tab:downstream_task}, an additional OML dataset has been used to investigate the SBP and DBP predictions in longitudinal tracking.  
Additionally, typical frontier discriminative methods including InceptionTime~\cite{inceptiontime}, ResNeXt-101~\cite{resnext} and CoAtNet~\cite{coatnet}, will be involved for the downstream task comparison. As we can see, our framework improves upon the state-of-the-art models remarkably. It is obvious that our proposed modules not only effectively work on the short-term waveform transformation task but also shed light on their potential to address the BP longitudinal tracking.
\begin{table}[!t]
    \renewcommand\arraystretch{1.35}
    \centering
    \scalebox{1.0}{
    \small
    \begin{tabularx}{1\linewidth}{@{}X c c c c c@{}}
    \toprule[1pt]
         \multirow{2}{*}{Methods}  & \multicolumn{2}{c}{SBP}  & \multicolumn{2}{c}{DBP} \\
     \cline{2-5}
       & MAE & ME±STD & MAE & ME±STD \\
        
        \midrule
        InceptionTime & {15.51} & {-4.70±17.48}& {9.10} & {-3.80±10.56}  \\
        ResNeXt-101 & {11.36} & {-1.72±12.48}& {8.31} & {-0.46±9.61} \\
        CoAtNet & {10.79} & {-1.61±11.26} & {7.02} & {0.76±8.05}\\
        SwinTransformer& {11.17} & {-1.08±11.83} & {7.35} & {0.79±9.03}\\
        PPG2ABP & {9.18} & {-1.74±9.50} & {7.29} & {0.36±7.80} \\
        Time-VQVAE & {8.92} & {-0.89±8.32} & {6.87} & {-0.48±7.02} \\
        VQ-VAE & {8.63} & {-1.37±8.35} & {6.73} & {-0.82±6.93} \\
        Ours & \textbf{8.04} & \textbf{-1.19±8.18} & \textbf{6.49} & \textbf{-0.72±6.75}\\
        \bottomrule
    \end{tabularx}}
    \captionsetup{width=1.\linewidth}
    \caption{Generalization of downstream task on the private OML dataset.}
    \label{tab:downstream_task}
\end{table}

\section{Conclusion}
This paper first speculates the latent space shift in anomalous PPG-to-ABP waveform transformation. Afterward, we propose a Latent Space Constraint Transformer (LSCT) to address this problem. Firstly, we utilize discretized codebook to quantize the latent space. Secondly, the Correlation-boosted Attention Module (CAM) has been proposed to correct the selection bias of codebook, which can substantially diminish the latent space shift from the anomalous inputs. Lastly, a Multi-Spectrum Enhancement Knowledge (MSEK) is introduced to present graph embedding in the channel view, which promotes the expressiveness of the latent code. Our method achieved competitive performance compared with the existing state-of-the-art waveform transformation methods. It also possesses substantial possibilities for the generalization of downstream tasks like BP longitudinal tracking.
In the future, more endeavors will be made to explore its efficacy across broader waveform transformation tasks.

\bibliography{aaai24}

\end{document}